# Insights into the Unusual Semiconducting Behavior in Low-Dimensional Boron

Shao-Gang Xu,[†,‡,§] Xiao-Tian Li,[‡] Yu-Jun Zhao,[‡] Wang-Ping Xu,[†] Ji-Hai Liao,[‡] Xiu-Wen Zhang,[§] Hu Xu[*,†] and Xiao-Bao Yang[*,‡]

[†]*Department of Physics, Southern University of Science and Technology, Shenzhen 518055, P. R. China* [‡]*Department of Physics, South China University of Technology, Guangzhou 510640, P. R. China*

[§]*College of Electronic Science and Technology, Shenzhen University, Shenzhen 518060, P. R. China*

## ABSTRACT

Elementary semiconductors are rare and attractive, especially for low-dimensional materials. Unfortunately, most of boron nanostructures were found to be metallic, despite of their typical semiconducting bulk structure. Herein, we propose a general recipe to realize low-dimensional semiconducting boron. This unusual semiconducting behavior is attributed to charge transfer and electron localization, induced by the symmetry breaking that divides boron atoms into cations and anions. In addition, it is feasible to accomplish band gap engineering by rationally designing various structures. Importantly, the low-dimensional semiconducting boron are predicted to be an excellent solar-cell material with the power conversion efficiency of higher than 20%, paving the way for their promising optoelectronic applications.

[*]E-mail address: xuh@sustc.edu.cn
[*]E-mail address: scxbyang@scut.edu.cn

The filling of electronic bands has generally distinguished metals and insulator at zero temperature, depending on the position of the Fermi level which lies in a band gap in insulators or inside a band for metals. In elementary bulk, metals possess high number of coordinates with the metallic characteristic due to the electron delocalization, while the number of coordinates in most elementary semiconductors is low except boron (B). The unique electron deficiency enables the formation of complicated bonding in bulk boron phases,[1] where the rhombohedral bulk is comprised of a highly symmetric icosahedral $B_{12}$ in a crystalline state.[1-2]

Following the upsurge of graphene, a diverse array of atomically thin materials have been proposed due to their outstanding properties,[3] such as the few-layer black phosphorus and transition metal dichalcogenides (TMDs), which can be designed as low-dimensional optoelectronic devices due to their semiconducting characteristic.[4-5] The experimental observation of planar clusters ($B_{35}^-$, $B_{36}$ .*et al*) are considered to be the precursor for borophene,[6-9] and boron monolayers have been achieved on the silver (Ag) substrates with hexagonal vacancies of various concentrations and distributions,[10-11] which have been confirmed by the first-principles calculations.[12-13] In contrary to the semiconducting bulk,[2, 14] all the two dimensional (2D) boron synthesized by experiments are metallic,[10-11] and the metallic features are due to the out-of-plane $p_z$-derived bands.[15-17] Up to now, a number of unique metallic characteristics have been proposed in various 2D boron allotropes, including the visible range plasmonics,[18] Dirac cones[19-20] and superconductivity.[21-22]

The metallic characteristic of low-dimensional boron is attributed to excess electrons induced by high number of coordinates, e.g. the high energy antibonding states are occupied in the triangular boron monolayer.[15] In order to transfer the excess electrons and decrease the electron delocalization, the symmetry breaking is necessary, including the introducing of the vacancy defect,[15-16, 23] forming the local covalent B-B bond, and chemical functionalization. In such a case, boron atoms might be divided into cations and anions, where the charge transfer will enhance the electron localization and consequently induce the semiconducting behavior.

To confirm the idea, we introduce covalent interlayer bonding between two triangular boron sheets, which will also enhance the stability of free-standing 2D boron allotropes.[24-25] Based on the average electron compensation (AEC) mechanism,[26] we show that there is electron excess of 1/3 $e$ for every boron atom in an ideal triangular sheet. For a super-cell with $3n$ ($n=1, 2, ...$) boron atoms, $2n$ delocalized electrons will become localized by forming $n$ covalent bonds between the upper and lower boron sheets. Herein, we reveal that the unusual semiconducting behavior in low-dimensional boron is attributed to the symmetry breaking, followed by the localization of $p_z$ orbitals and the charge transfer. All the first-principles calculations were based on the density functional theory (DFT) implemented in the Vienna *ab initio* simulation package (VASP) method.[27] The electron-ion interactions were described by the projector augmented wave (PAW) potentials.[28] To treat the exchange-correlation interaction of electrons, we chose the Perdew-Burke-Ernzerhof (PBE) functional within the generalized-gradient approximation (GGA).[29] In addition, the Heyd-Scuseria-Ernzerhof (HSE06) hybrid functional[30] and the quasi-particle $G_0W_0$ approach[31] were applied for the band gap corrections, and the GW–Bethe-Salpeter equation (GW-BSE) approach[32-35] was included to investigate the excitonic properties of the 2D semiconductors. More computational details are presented in the Supporting Information (SI).

Taking $p(2\times3)$ and $p(3\times3)$ bilayers as examples, we have constructed two typical 2D boron allotropes (denoted by R1 and R2) as shown in Figure 1, with two/three interlayer bonds in R1/R2 respectively. As shown in Figure 1(c, d), the HSE06 band structure calculations confirmed that R1 and R2 phases are semiconductors with the indirect band gaps of 0.305 eV and 1.428 eV, respectively. The projected band structure reveals that the valence band maximum (VBM) is dominated by the in-plane ($s+p_{x,y}$) orbitals and the out-of-plane ($p_z$) orbitals, while the conduction band minimum (CBM) is mainly from the in-plane ($s+p_{x,y}$) orbitals. The relaxed structural parameters of R1 and R2 are listed in Table S1. In addition, the phonon dispersions along all the high-symmetry lines and molecular dynamics snapshots at a higher temperature (see Figure S1) indicate that the two new phases have high dynamic and thermal stabilities.

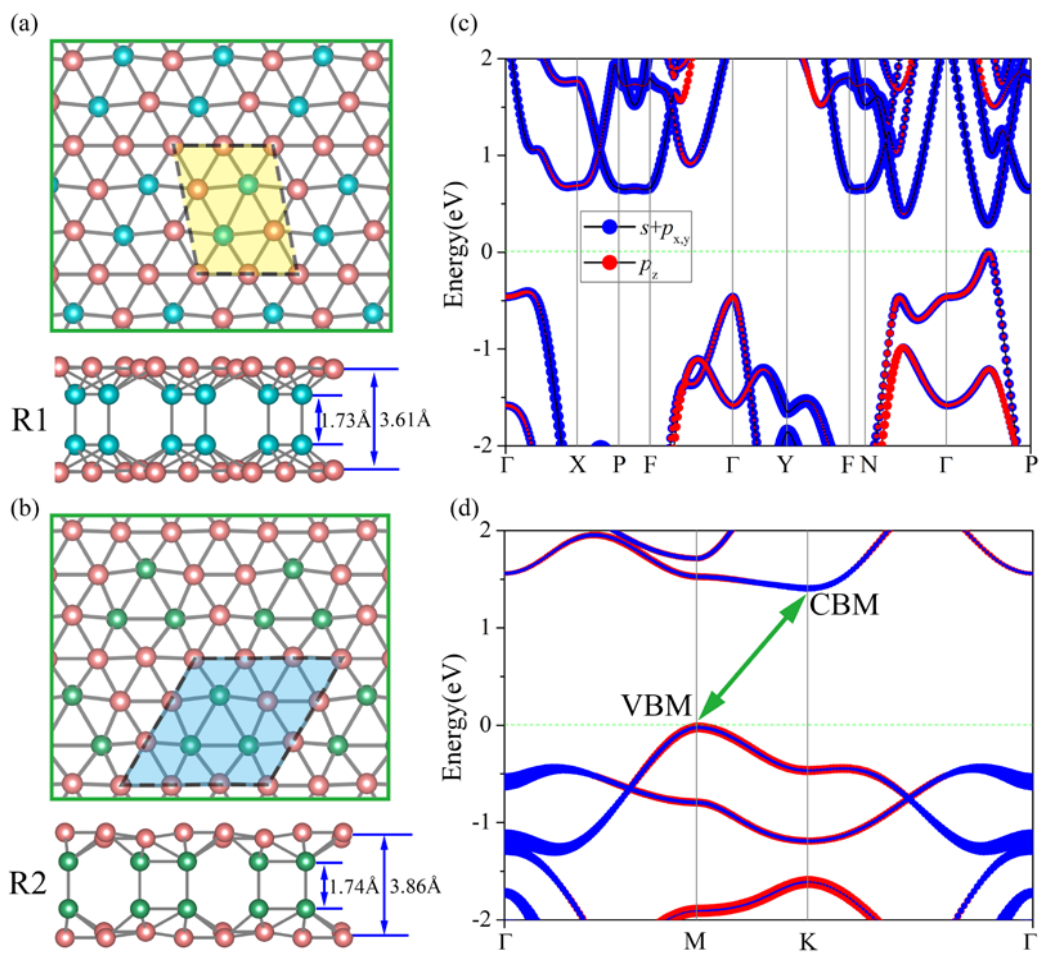

Figure 1. (a, b) Atomic structures of the new sandwiches R1 and R2. The black dash lines represent the unit cells of the corresponding structures. The cyan and green atoms in figure (a, b) represent the boron atoms for the interlayer bonds. (c, d) The projected HSE06 band structures of R1 and R2. (VBM is set to 0.)

In order to explore the possibility of boron sandwiches growth on the metal substrates, we have shown the thermodynamic phase diagram of the B/metal systems in Figure 2. As we know, several boron monolayers have been fabricated on the Ag(111) substrates by the molecular beam epitaxy (MBE) method, including $\beta_{12}$, $\chi_3$[11] and the metastable α sheet.[36] Clearly, the phase diagrams indicated that the monolayer boron phases on various metal substrates are more stable than the boron films at B-poor conditions. However, the synthesis of atomically thin $\gamma_{28}$ film on copper foils has been realized by chemical vapor deposition (CVD) method.[37] The experimental works observed the deposited boron would diffuse into the catalytic Cu surface, and since the system reaches super-saturation, the boron would nucleate as thin film to counteract the continuous feeding of B onto the substrates.[37] The formation energies in the phase diagrams indicated that the B-rich conditions will stabilize the boron films on the metal substrates, since the semiconducting R1 sandwich is more stable than the experimental $\gamma_{28}$ film on Cu(111) substrates, revealling the possibility to fabricate the 2D semiconducting boron allotropes by the control of growth conditions. Notably, the absorption spectroscopy analysis demonstrated the semiconducting property of the experimental boron film,[37] while the proposed $\gamma_{28}$ film is found to be metallic.[38] Since the R1 sandwich is more stable than the $\gamma_{28}$ film on the metal substrates, we are inclined to believe that the new semiconducting phases proposed by us are close to the phases detected by the experiments.

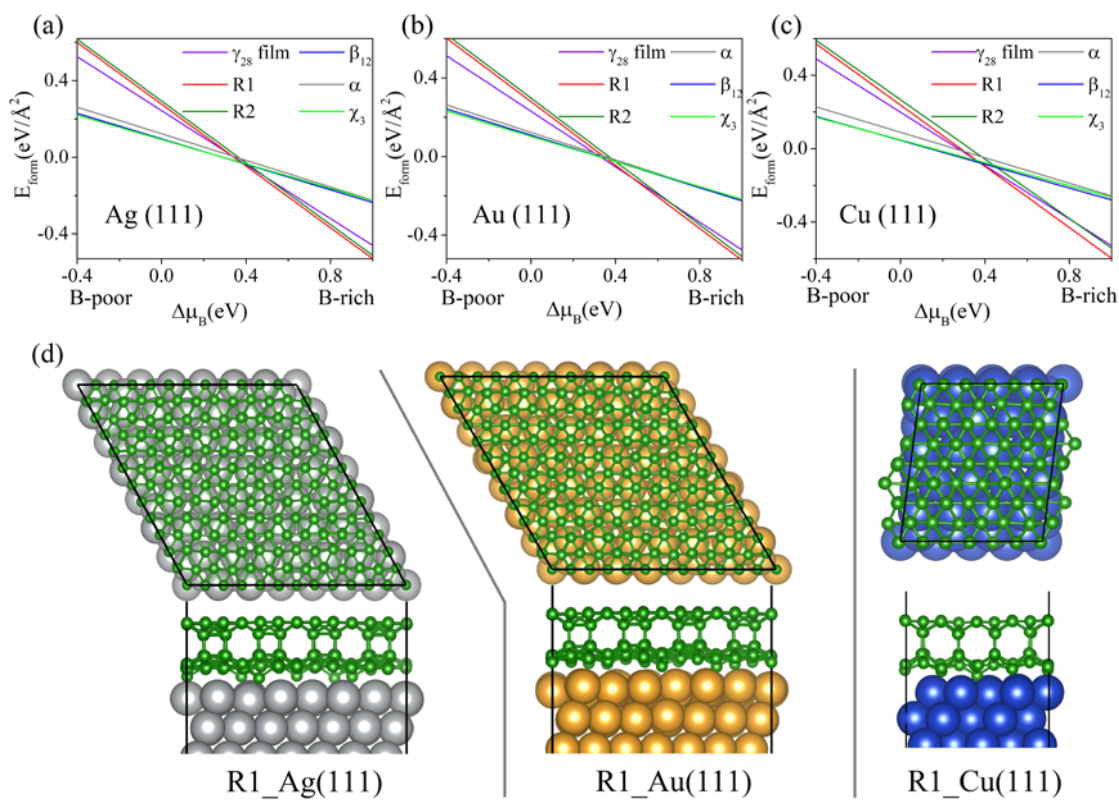

Figure 2. (a-c) represent the formation energies($E_{form}$) of various 2D boron allotropes on metal substrates as a function of B chemical potential ($\Delta \mu_B = \mu_B - \mu_{B(crystal)}$). (d) The atomic structures(top view and side view) of the R1 sandwich growth on the metal substrates

As shown in Figure S2, we analyze the features of interlayer B-B bonds in R1 and R2 by calculating its electron localization function (ELF), demonstrating the electron localization around B-B bonds at the top(bottom) planes and the interlayer $B_4/B_6$ units. The 2D ELF plots for the sliced planes in Figure S2 also indicate the strong covalent electrons in the interlayer B-B bonds, in agreement with the Bader charge analysis showing charge transfer of $0.55e(0.87e)$ from the $B_4$-rectangle($B_6$-trigonal-prism) to the neighboring boron planes. The above mechanism is analogous with the former 2D sandwich metal borides,[39] where the boron planes and metal plane act as the "anions" and "cations", respectively. Based on the structure recognition algorithm,[40] we have performed a full screening of possible 2D boron allotropes using $p(2\times3)$ and $p(3\times3)$ super-cells with 2/3 interlayer bonds. The results of high-throughput electronic properties screening revealed that R1 and R2 are unique semiconductors, implying that the distribution and concentration of the interlayer bonds codetermine the presence of semiconducting properties. Note that the localization of metallic $p_z$ orbitals is attributed to the interlayer B-B bonds according to the projected density of states (PDOS) analysis (see Figure S3), where there is a sharp increase of $p_z$ electrons from the boron planes to the $B_6$ units. However, the PDOS analysis fails to explain the semiconductor behavior, since the distributions of $p_z$ orbitals are almost the same for semiconducting phase(R2) and metallic phase(M2) as shown in Figure S3.

Next, we have compared the charge distribution of the R2 isomers with three interlayer bonds (shown in Figure 3). The Bader charge analysis shows that there is charge dissipation in the interlayer $B_6$ triangular prism of R2, the "cations" part, and the charge will be accepted by the $B_3$ triangle on the upper/lower planes, the "anions" part. There is no charge transfer in the other $B_3$ triangle, e.g. "neutral" part. As shown in Figure 3(a), the "cations" and "anions" triangles form a honeycomb lattice with the "neutral" triangle trapped in the center of hexagon. The large gap of R2 is attributed to the charge transfer from the "cations" triangles to the "anions" ones, similar to the famous boron nitride monolayer. For the metallic M2 in Figure 3(b), the charge transfer is mainly from the two B atoms and it is accepted by one B atom, the "neutral" B atoms form a connected network and induce the metallic characteristic. Similarly, the "neutral" B atoms in metallic M3(Figure 3c) form a periodic chain, besides the "cations" and "anions" triangles. Therefore, the connected network formed by the "neutral" B atoms may act as a delocalized metallic channel, and the semiconducting behavior is attributed to the symmetry breaking and charge localization.

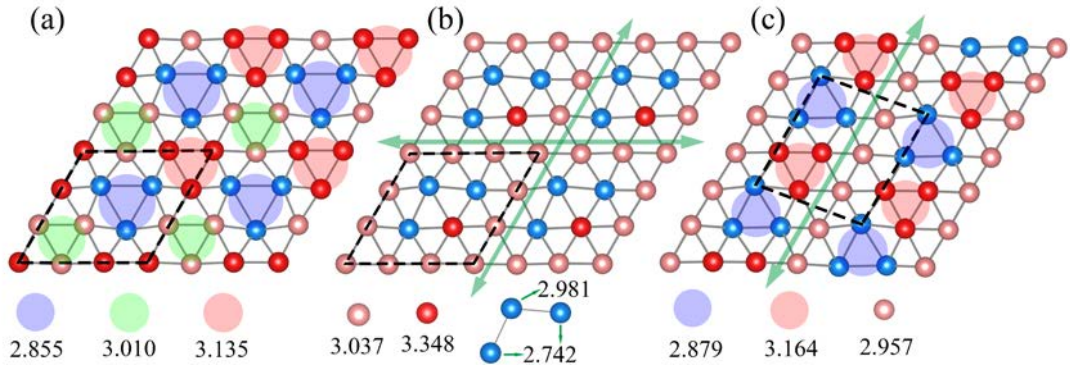

Figure 3. The results of Bader charge analysis for the *p*(3×3) super-cell boron allotropes (a-R2, b-M2 and c-M3). The values present here are the average value of the given color regions(a, c) or the color atoms(b, c). The blue atoms represent the boron atoms for the interlayer bonds.

To have a more comprehensive view of the effect of covalent bond, we have considered the surface decoration with hydrogen (H) adsorption in a boron monolayer, where the B-H bond of adsorption site is a strong covalent bond. Following the idea of AEC mechanism, we have tested 2~5 H atoms adsorption for the 6~15 times super-cells (see Figure S4). The band structures in Figure S5 have revealed that all these hydrogen adsorption systems are semiconductors. The strong B-H interactions lead to the redistribution of electron density, which can be confirmed by the obvious charge transfer of B/H atoms. For the RH-2 phase in Figure 4(a, b), there is charge transfer of 0.524 *e* from each B atom to H atom at the adsorption site, and therefore the B atoms can be divided into three parts ("cations", "anions" and "neutral"), which is in agreement with the charge distribution in R2 sandwich. For the RH-3 phase in Figure 4(c, d), there are two kinds of H atoms in the "Y"-like adsorption configuration, and the center H atom obtain fewer electrons from the B atom. The inequivalent adsorption sites of RH-3 lead to the high deformation of the B triangular plane, and there are five kinds of B atoms based on the charge transfer analysis. Similarly, both the "neutral" regions in above two semiconducting 2D B-H systems are disconnected by the charged regions, indicating the universality of semiconducting behavior in low-dimensional boron.

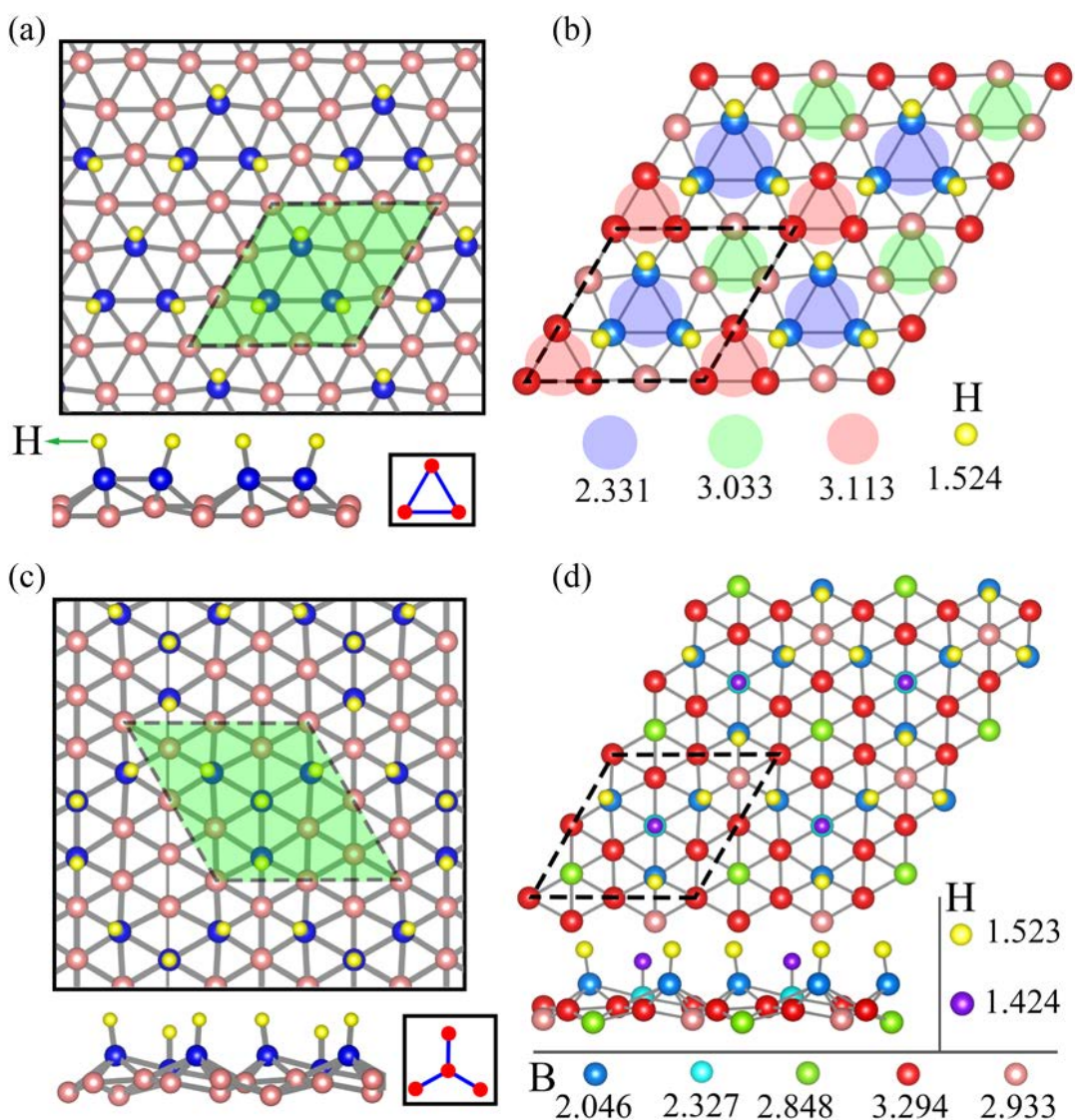

Figure 4. (a, c) The atomic configurations of hydrogen adsorption on $p(3\times3)$ and $p(2\sqrt{3}\times2\sqrt{3})$ super-cells. The blue atoms represent the boron atoms at the adsorption sites. The black dash lines represent the unit cells of the corresponding structures. (b, d) The results of Bader charge analysis for the 2D B-H systems. The values present here are the average value of the given color regions(b) or the color atoms(d).

In the following, we focus on the possible band gap engineering with low-dimensional boron and the potential applications. With the high-throughput screening, we have found two new phases (R3, R4) are semiconductors with indirect band gap of 0.862 eV and 0.975 eV, respectively (shown in Figure S6(a)). Interestingly, R3 and R4 can be viewed as the derivative products of R1 and R2, indicating a general rule to design the 2D semiconducting boron allotropes based on the two basic unit cells of R1($T_A$) and R2($T_B$). As shown in Figure S6(b), various 2D boron allotropes will be obtained by modulating the numbers and sequences of the two parts (A,B) in the super-cell($T_{m \times A + n \times B}$). The hybrid functional band structure calculations have confirmed that all the designed sandwiches are semiconductors(shown in Figure S8), and the band gaps of new phases are located in the range of $T_A$(0.305 eV) and $T_B$(1.428 eV). As shown in Figure S6(d), the band gap of the corresponding structure is increased with the increasing ratio ($n/m+n$) of $T_B$ in the super-cell. Notably, the structures of the same ratio may result in various band gaps, as the band gap of $T_{ABAAB}$ is much larger than $T_{AABBA}$. Thus, both the ratio of A/B and the orders of basic parts (A, B) in the super-cell will modulate the band gaps of these low-dimensional boron structures. As shown in Table S1, the average formation energy($E_{form}$) difference of the new sandwiches are smaller than 0.03 eV/atom, which due to the close stability of $T_A$ and $T_B$. All of the new phases are stable than the boron monolayers[16] and $P6/mmm$,[25] and R3 is more stable than the semimetallic $Pmmn$.[24]

For potential application in real systems, 2D semiconductors should be grown on a flexible substrate, where the strain effect would inevitably be considered due to the lattice mismatch. As shown in Figure 5(a), the HSE06 band gap of R2 would be effectively modulated with the biaxial strain from 1.23 to 1.58 eV. Notably, the R2 sandwich would transform to a direct gap semiconductor under the 2% compression strain. In order to obtain the accurate quasi-particle (QP) band gap ($E_g$), we have calculated the $G_0W_0$ band structure of compressed(-2%) R2, showing its fundamental band gap of 1.48 eV in Figure 5(b). The inset of Figure 5(b) shows the calculated $G_0W_0$-BSE spectrum of compressed(-2%) R2, and the optical gap is 1.16 eV, indicated by the first peak of the absorption spectrum. The corresponding exciton binding energy ($E_b$) obtained here is 0.32 eV, which is satisfied with the linear scaling law ($E_b \approx E_g/4$) of 2D semiconductors.[41]

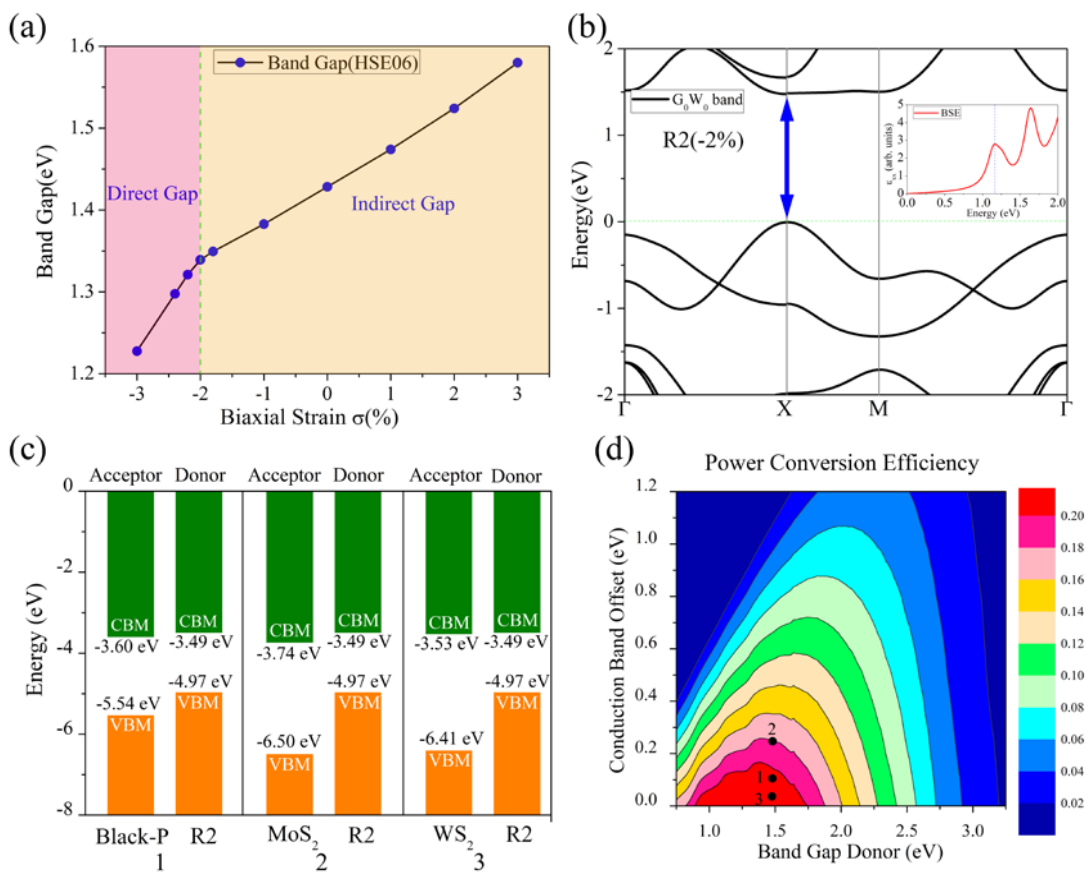

Figure 5. (a) The relationship of HSE06 band gaps of R2 and biaxial strain. (b) The $G_0W_0$ band structure of compressed R2 (VBM is set to 0), and the inset represents the BSE optical absorption spectrum of the corresponding structure. (c) Band offsets between black-P/$MoS_2$/$WS_2$ (acceptor) and the compressed R2 (donor). (The surrounding vacuum is the reference energy.) (d) Computed power-conversion efficiency contour as a function of the donor band-gap and conduction band offset.

It is well known that semiconductors with direct gap around 1.4 eV are ideal candidates for the solar absorption material. The band-edge energies VBM and CBM of compressed(-2%) R2 are -4.97 eV and -3.49 eV, respectively. Matching band alignment with appropriate semiconducting acceptors, the direct-gap R2 may be applied as solar cell donor material. Based on the band-gap-center approximation,[42] we have checked the QP band-edge energies of many 2D semiconductors, and found three potential acceptors(black-P, $MoS_2$ and $WS_2$) to form a type-II heterojunction alignment with compressed(-2%) R2 (Figure 5c). The definition of power conversion efficiency (PCE) η is provided in the SI. As shown in Figure 5(d), solar systems constructed with $WS_2$/R2 and black-P/R2 can achieve PCEs as high as ~22%/20.7%, and the PCE of $MoS_2$/R2 is 18.1%. These values are comparable to proposed bilayer phosphorene/$MoS_2$ heterojunctions (16−18%),[43] hybrid PCBM/CBN heterojunctions (10−20%)[44] and g-$SiC_2$ based systems (12−20%) for highly efficient solar cells.[45] Notably, for the 2D B-H systems, there are many semiconductors with direct band gap, for example the hybrid functional band gap of RH-3a in Figure S(4, 5) is 1.18 eV, indicating their potential applications for advanced optoelectronic devices.

In summary, we have theoretically predicted 2D semiconducting boron allotropes by effectively introducing the symmetry breaking and charge transfer in triangular boron sheet. The semiconducting behavior is attributed to the charge transfer between boron atoms from different region, where the neutral region should be disconnected. This universal mechanism is also verified by the metallic triangular boron sheets with hydrogen adsorption. By modulating the ratio and orders of the two basic parts(R1, R2) in the unit cell, various boron sandwiches are designed to realize the band-gap engineering. The predicted PCE for $WS_2$/R2 heterojunctions can be as high as 22%, rendering the solar systems an efficient candidate in flexible optoelectronic device. We are inclined to believe that the new type 2D boron allotropes may be synthesized on the metal substrates at B-rich conditions, and the proposed efficient 2D boron-based solar cell may be achieved by experiments. The discovery of semiconducting phases provides more future opportunities for 2D boron allotropes research.

This work was supported by the National Natural Science Foundation of China (No. 11474100, 11674148, 11574088), the Guangdong Natural Science Funds for Distinguished Young Scholars (No. 2014A030306024, 2017B030306008), and the Guangdong Natural Science Funds for Doctoral Program (Grant No. 2017A030310086). The computer times at the National Supercomputing Center in Guangzhou (NSCCGZ) and Guangzhou Ginpie Technology are gratefully acknowledged.

S.G.X. and X.T.L contributed equally to this work.

# Supporting Information

# Insights into the Unusual Semiconducting Behavior in Low-Dimensional Boron

*Shao-Gang Xu[+], Xiao-Tian Li[+], Yu-Jun Zhao, Wang-Ping Xu, Ji-Hai Liao, Xiu-Wen Zhang, Hu Xu[*] and Xiao-Bao Yang[*]*

[*]Corresponding author Email: xuh@sustc.edu.cn; scxbyang@scut.edu.cn

## The Computational Details:

All structures are fully relaxed until the force on each atom is smaller than -0.01 eV/Å with the cutoff energy of 480 eV. To avoid the interaction between adjacent images, a vacuum region of 21 Å along the *z* direction was added for the model. To confirm the dynamical stability, the phonon band structures were calculated with the finite displacement method as implemented in the Phonopy program.[1] Thermal stability was also studied using *ab initio* molecular dynamics (AIMD) simulations with the temperature controlled by a Nosé heat bath scheme.[2]

## Average Formation Energy:

The average formation energy ($E_{form}$) of the B sandwiches is defined as $E_{form} = (E_{tot}/n - E_{at})$, where $E_{tot}$ is the total energy of the B sandwiches, $E_{at}$ is the energy of an isolated spin-polarized boron atom, and *n* is the number of B atoms in the cell.

## Thermodynamic Phase Diagram:

In order to further examine the stability of different 2D B allotropes deposited on metal substrates, we define the average formation energy($E_{form}$) as

$$E_{form} = \frac{1}{A}(E_{tot} - E_{sub} - n \times \mu_B)$$

where the $E_{tot}$, $E_{sub}$ represent the total energy of the B/metal systems and the metal substrates respectively, *n* is the number of B atoms of the absorbed 2D boron allotropes, *A* is the area of the 2D boron structures, and $\mu_B$ is the chemical potential of B atoms. Under the B-rich condition, $\mu_B$ is referred to the per atom energy of α phase $\mu_{B(crystal)}$. Based on the above equation, the formation energies of various 2D boron structures on metal substrates as a function of B chemical potential ($\Delta\mu_B=\mu_B-\mu_{B(crystal)}$) can be obtained. The atomic structures of the 2D boron allotropes growth on the metal substrates are shown in Figure S9. Note that, we modeled the substrate using the five-layer slab with the bottom two layers fixed, considering a few possible configurations of 2D boron allotropes on the substrates where the absolute value of lattice mismatch δ is under 2.5%.

## Details of $G_0W_0$ Calculations:

In our $G_0W_0$ calculations on compressed(-2%) R2, we use a 9×9×1 k-mesh for the Brillouin zone sampling and 261 empty bands (There are 27 occupied bands). The cutoff energy for the plane wave basis set to be 450 eV, and the ENCUTGW is set to be 300 eV. Our test calculations show that the above parameters make sure that the computed band gap has an accuracy of 0.1 eV. To plot the quasi-particle band structure, we use the approach of maximally localized Wannier functions.[3]

## Power Conversion Efficiency:

The upper limit of the power conversion efficiency (PCE) $\eta$ is estimated in the limit of 100% external quantum efficiency (EQE)[4] with the formula given by

$$\eta = \frac{J_{sc}V_{oc}\beta_{FF}}{P_{solar}} = \frac{0.65\left(E_g^d - \Delta E_c - 0.3\right)\int_{E_g}^{\infty}\frac{P(\hbar\varpi)}{\hbar\varpi}d(\hbar\varpi)}{\int_0^{\infty}P(\hbar\varpi)\,d(\hbar\varpi)}$$

where the band-fill factor (FF) is taken to be 0.65, $P(\hbar\varpi)$ is taken to be the AM1.5 solar energy flux(expressed in W/m$^{-2}$/eV$^{-1}$) at the photo energy $\hbar\varpi$, and $E_g^d$ is the bandgap of the donor, and the $\left(E_g^d - \Delta E_c - 0.3\right)$ term is an estimation of the maximum open circuit voltage $V_{oc}$. The integral in the numerator is the short circuit current $J_{sc}$ in the limit of 100% EQE, and the integral in the denominator is the AM1.5 solar flux.

## Number of Electrons:

We have calculated the projected density of states (PDOS) for the 2D boron allotropes, and the number($N$) of electrons of the ($s$, $p_x$, $p_y$ and $p_z$) orbitals for the given B atom is defined as

$$N = \int_{-\infty}^{E_F} D(E)\,d(E)$$

where $E_F$ represents the Fermi level of the system, and $D(E)$ is the PDOS value for the given B

atoms.

Table S1. The symmetric groups(SG), calculated lattice constants by PBE functional, the band gaps calculated by HSE06 functional, and the average formation energy $E_{form}$ of all the 2D boron allotropes from GGA(PBE) results.

| Phase | SG | $a$(Å) | $b$(Å) | $\gamma$(°) | Bandgap(eV)-HSE06 | $E_{form}$(eV/atom)-PBE |
|---|---|---|---|---|---|---|
| $\eta_{1/8}$ | *Pmma*(51) | 5.07 | 11.709 | 90 | | -5.997 |
| *P6/mmm* | *P6/mmm*(191) | 2.865 | 2.865 | 120 | | -6.002 |
| *Pmmn* | *Pmmn*(59) | 4.520 | 3.260 | 90 | | -6.056 |
| *Pmmm* | *Pmmm*(47) | 2.88 | 3.26 | 90 | | -6.082 |
| $\gamma_{28}$-film | *P2/m*(10) | 5.624 | 6.988 | 90 | | -5.896 |
| R1($T_A$) | *P2/m*(10) | 3.332 | 4.408 | 79.11 | 0.305 | -6.043 |
| R2($T_B$) | *P-6m2*(187) | 4.986 | 4.986 | 120 | 1.428 | -6.027 |
| R3 | *Pbam*(55) | 3.333 | 8.659 | 90 | 0.861 | -6.059 |
| R4($T_{AB}$) | *Pm*(6) | 4.992 | 7.253 | 83.41 | 0.975 | -6.020 |
| $T_{AAB}$ | *Pm*(6) | 4.992 | 10.122 | 85.29 | 1.014 | -6.027 |
| $T_{ABB}$ | *Pm*(6) | 4.992 | 11.648 | 81.79 | 1.073 | -6.021 |
| $T_{ABBB}$ | *Pm*(6) | 4.992 | 15.874 | 86.99 | 1.050 | -6.022 |
| $T_{AAAB}$ | *Pm*(6) | 4.992 | 13.208 | 79.11 | 0.822 | -6.031 |
| $T_{ABBA}$ | *Pm*(6) | 4.992 | 14.507 | 83.41 | 0.808 | -6.026 |
| $T_{AAAAB}$ | *Pm*(6) | 4.992 | 15.874 | 86.99 | 0.632 | -6.033 |
| $T_{ABAAB}$ | *Pm*(6) | 4.992 | 17.293 | 90 | 0.994 | -6.024 |
| $T_{AABBA}$ | *Pm*(6) | 4.992 | 17.293 | 90 | 0.541 | -6.028 |

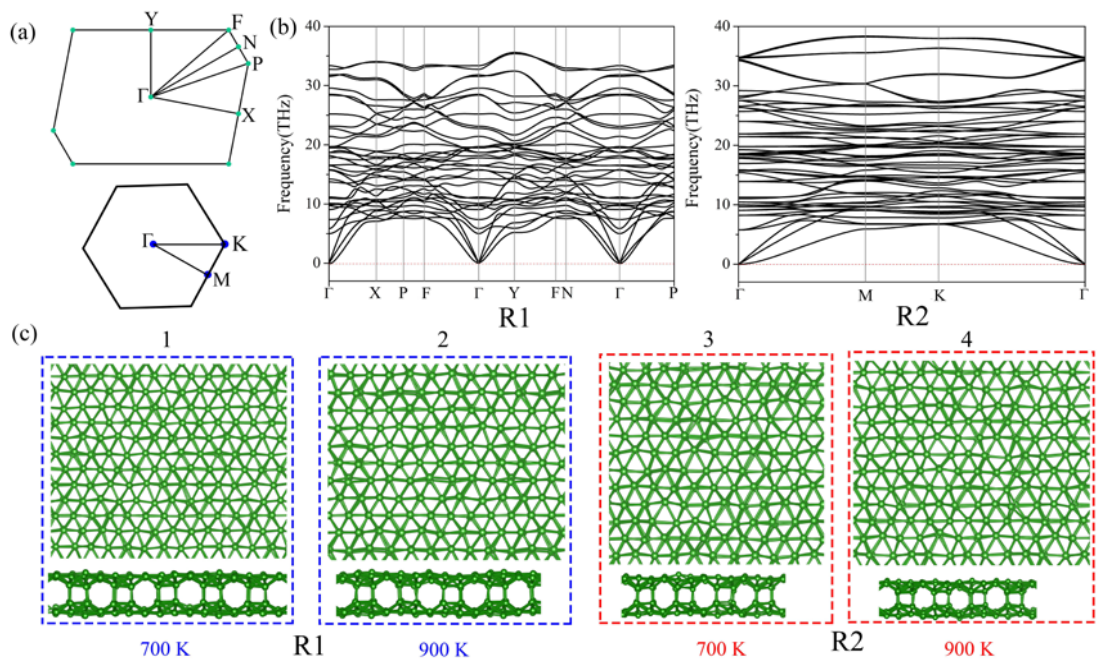

Figure S1. (a) The first Brillouin zones for R1 and R2. (b) The phonon dispersions along the high-symmetry line of R1 and R2. (c) The AIMD snapshots at the temperature of 700 K/900 K (10 ps) for R1 and R2, respectively.

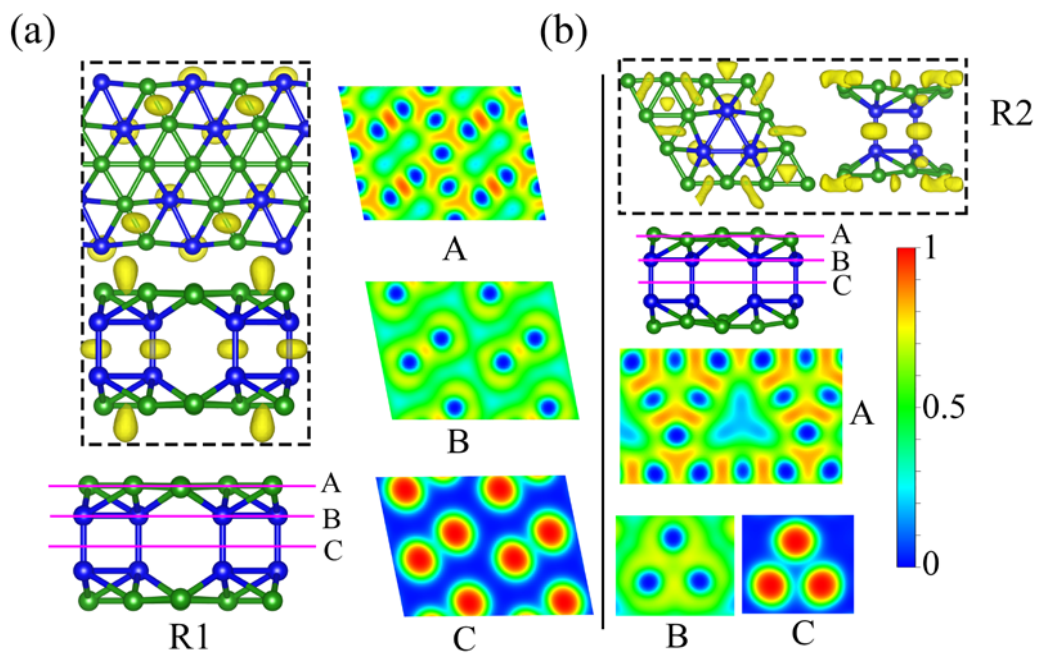

Figure S2. (a, b) The top view and side view of the electron localization function(ELF) of R1(0.85) and R2(0.82), and A-C display the 2D ELF of R1/R2 sliced planes.

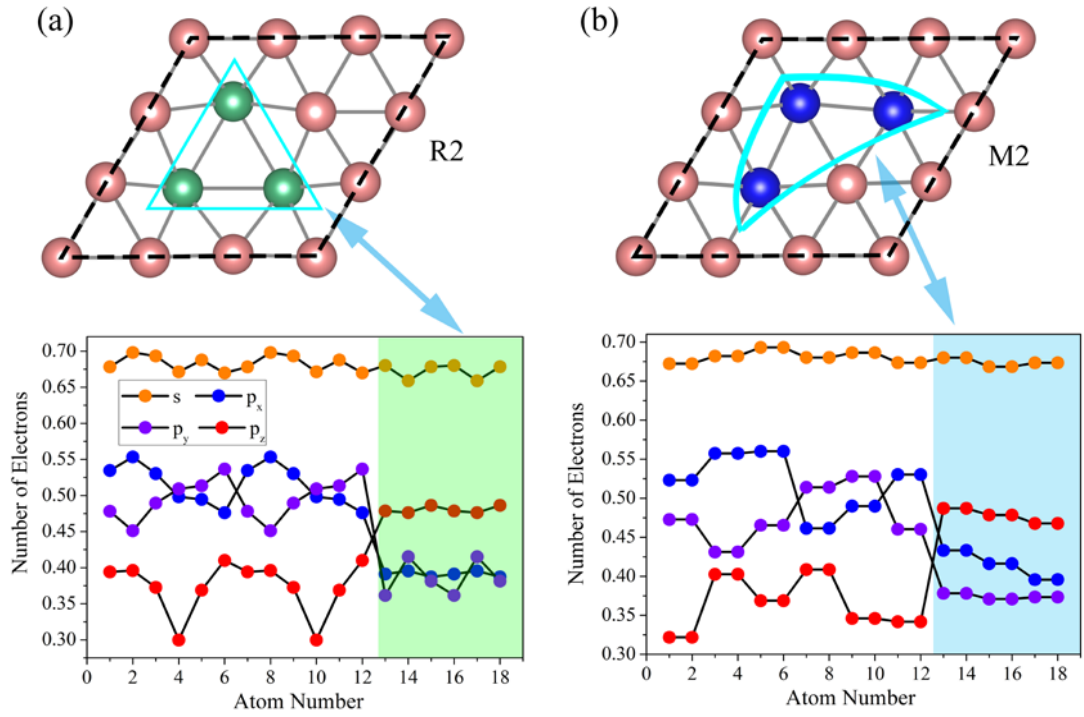

Figure S3. (a, b) The top view for the 2D boron allotropes(R2, M2), and the number of electrons of the projected orbitals for the given boron atoms. The green region and the blue region represent the results for the six boron atoms of the interlayer bonds.

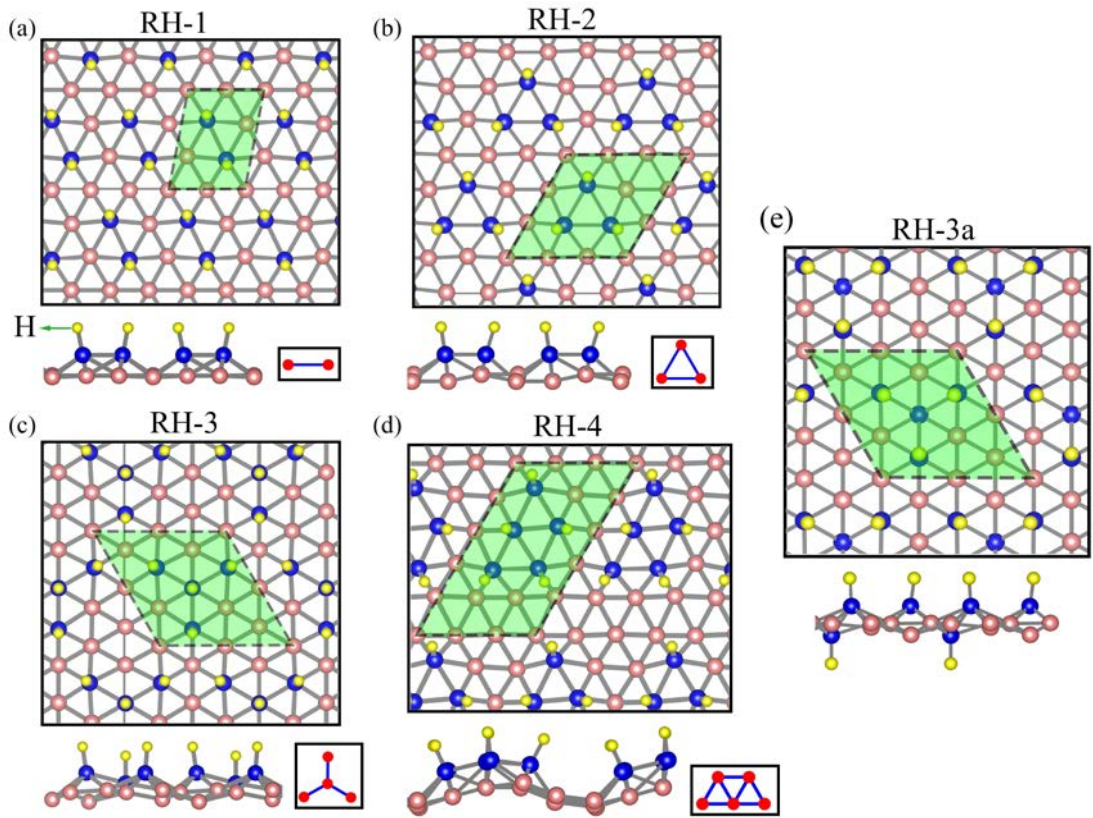

Figure S4. The atomic configurations of hydrogen adsorption on various triangular boron sheet super-cells. The blue atoms represent the boron atoms at the adsorption sites. The adsorption configurations are shown in the bottom corner of the corresponding structures. The black dash lines represent the unit cells of the corresponding structures.

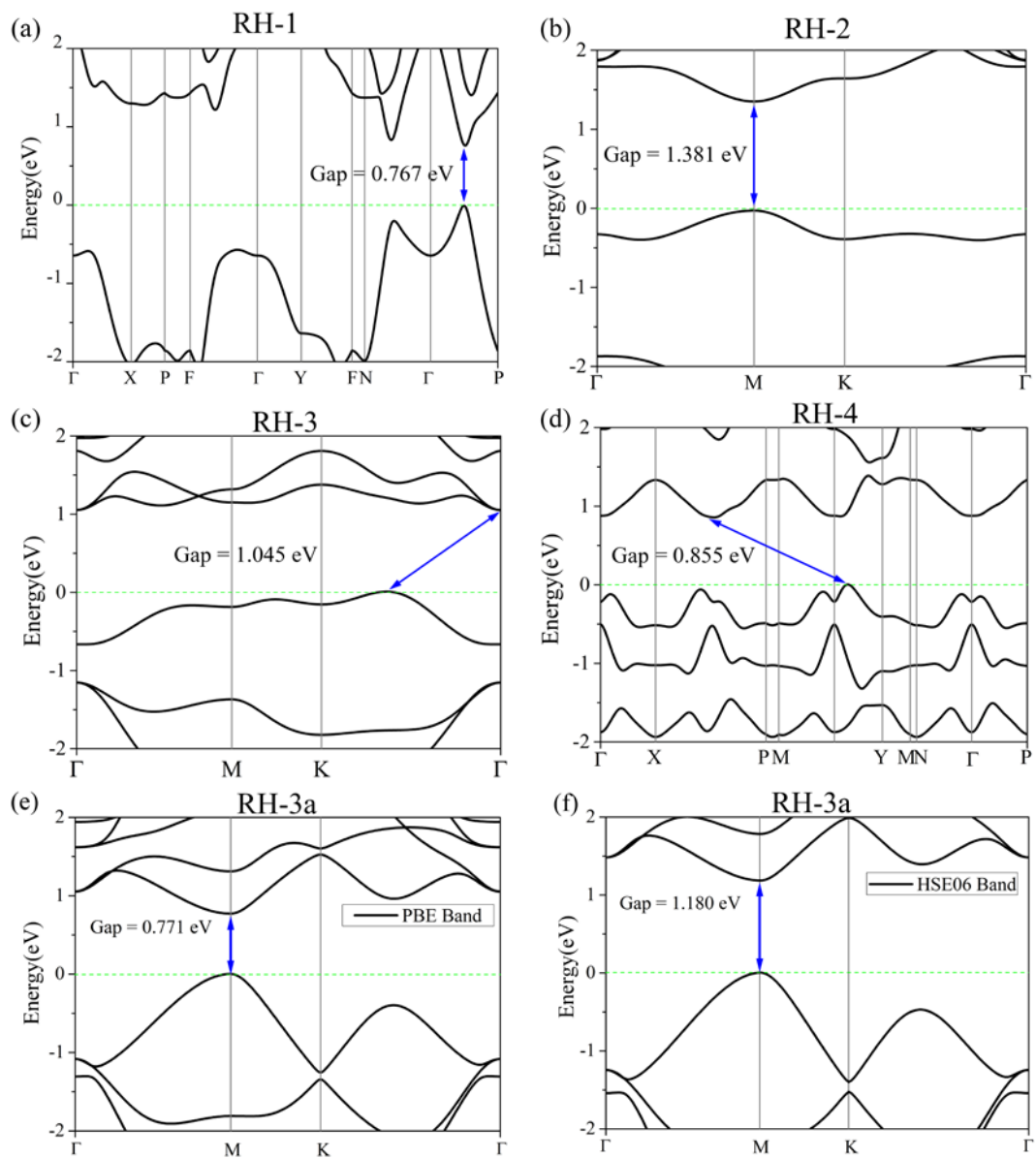

Figure S5. (a-d) The band structures of the hydrogen adsorption triangular boron sheet super-cells at PBE level. (e, f) The PBE and HSE06 band structures of 2D RH-3a phase. The blue arrows represent the position of the VBM and CBM of the corresponding adsorption systems (VBM is set to 0).

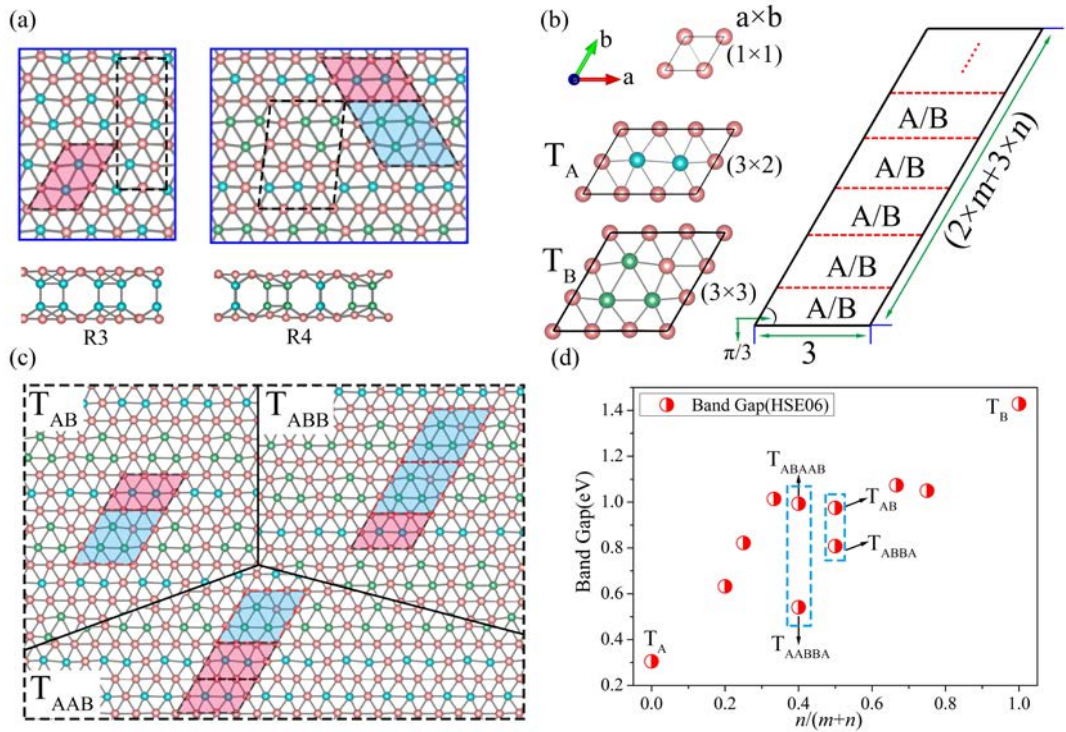

Figure S6. (a) The atomic structure of R3 and R4 (top view and side view). (b) The schematic diagram of constructing new boron sandwiches. (c) The top view of the combination products ($T_{AB}$, $T_{ABB}$ and $T_{AAB}$). (d) Relationship of band gaps and the ratio of part $T_B$ for the new 2D boron semiconductors.

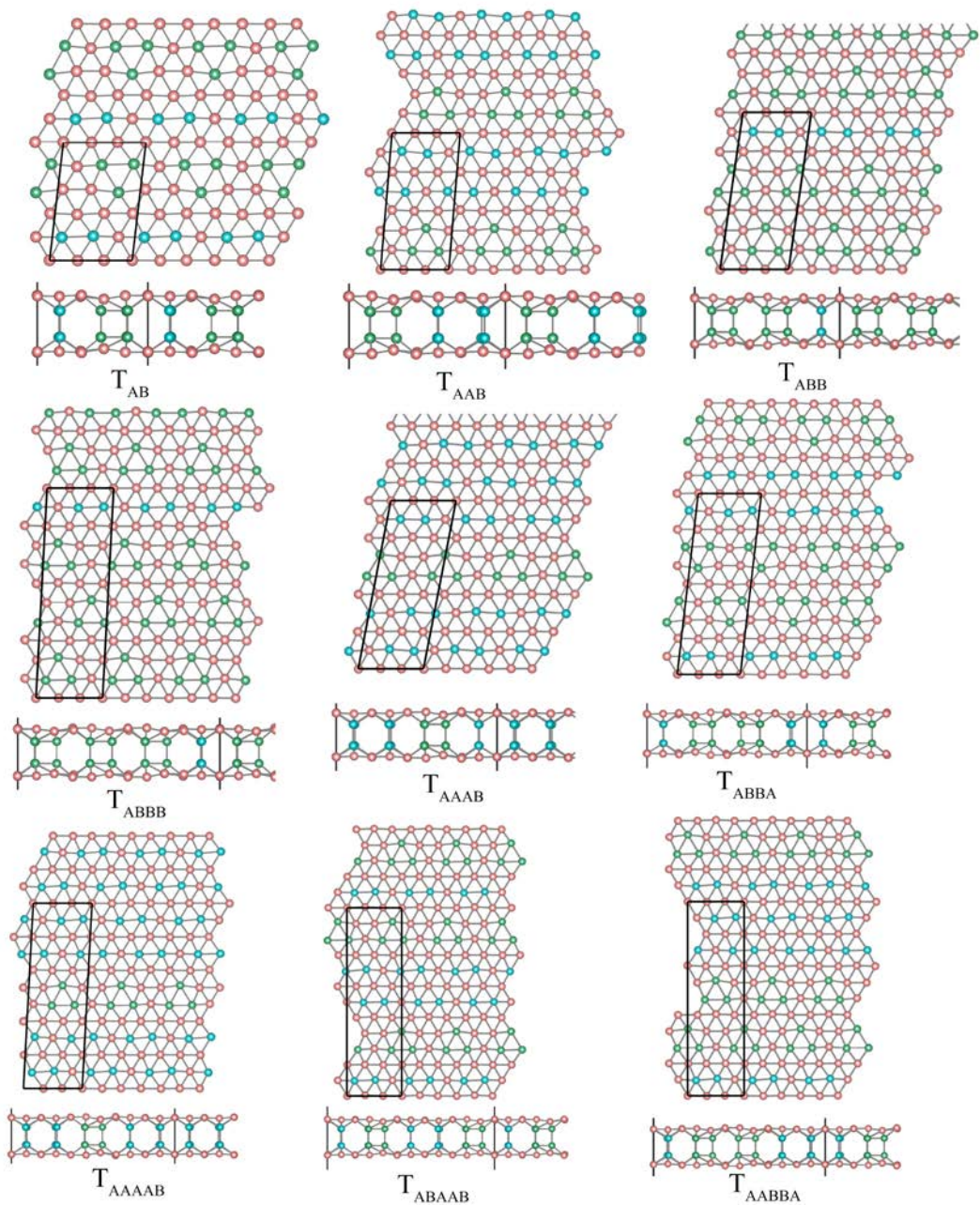

Figure S7. The atomic structures of the new designed stable boron sandwiches (top view and side view). The black solid lines represent the unit cells of the corresponding structures.

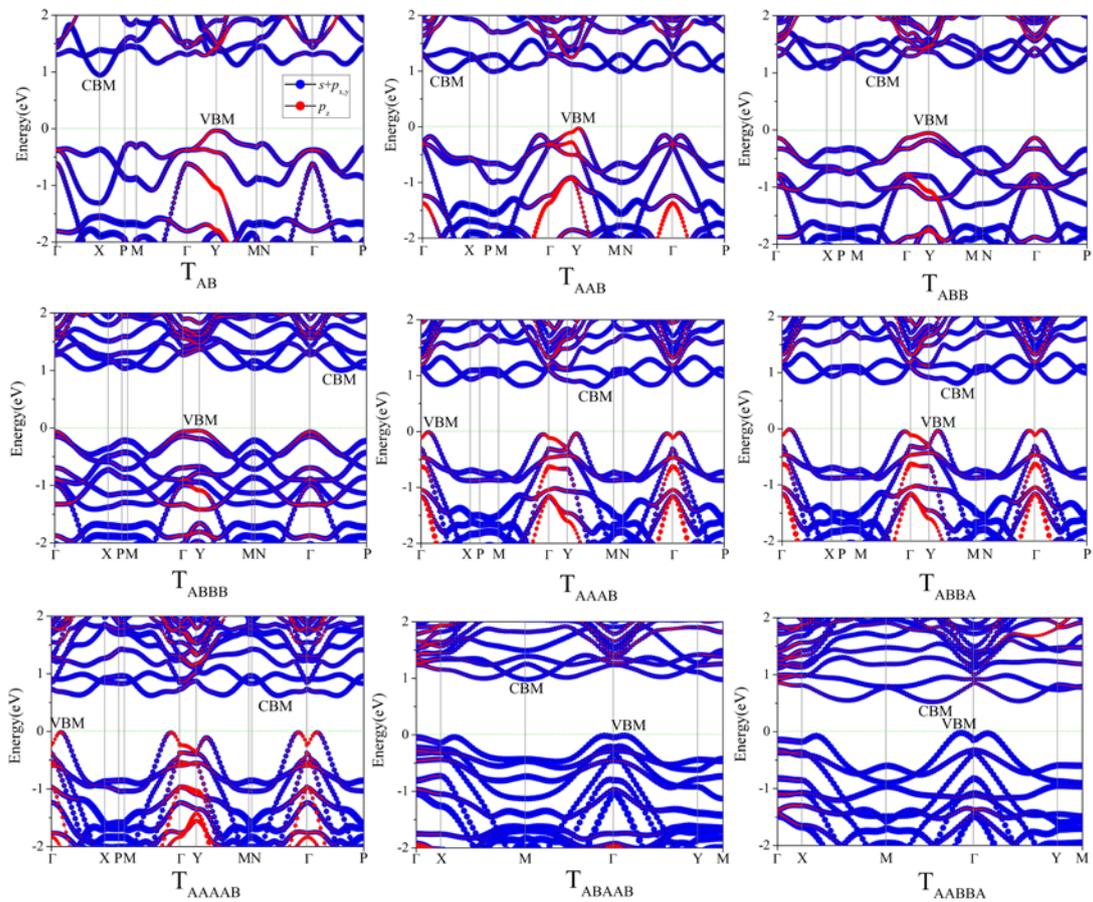

Figure S8. The projected band structures (HSE06) of the new designed boron sandwiches. (VBM is set to 0.)

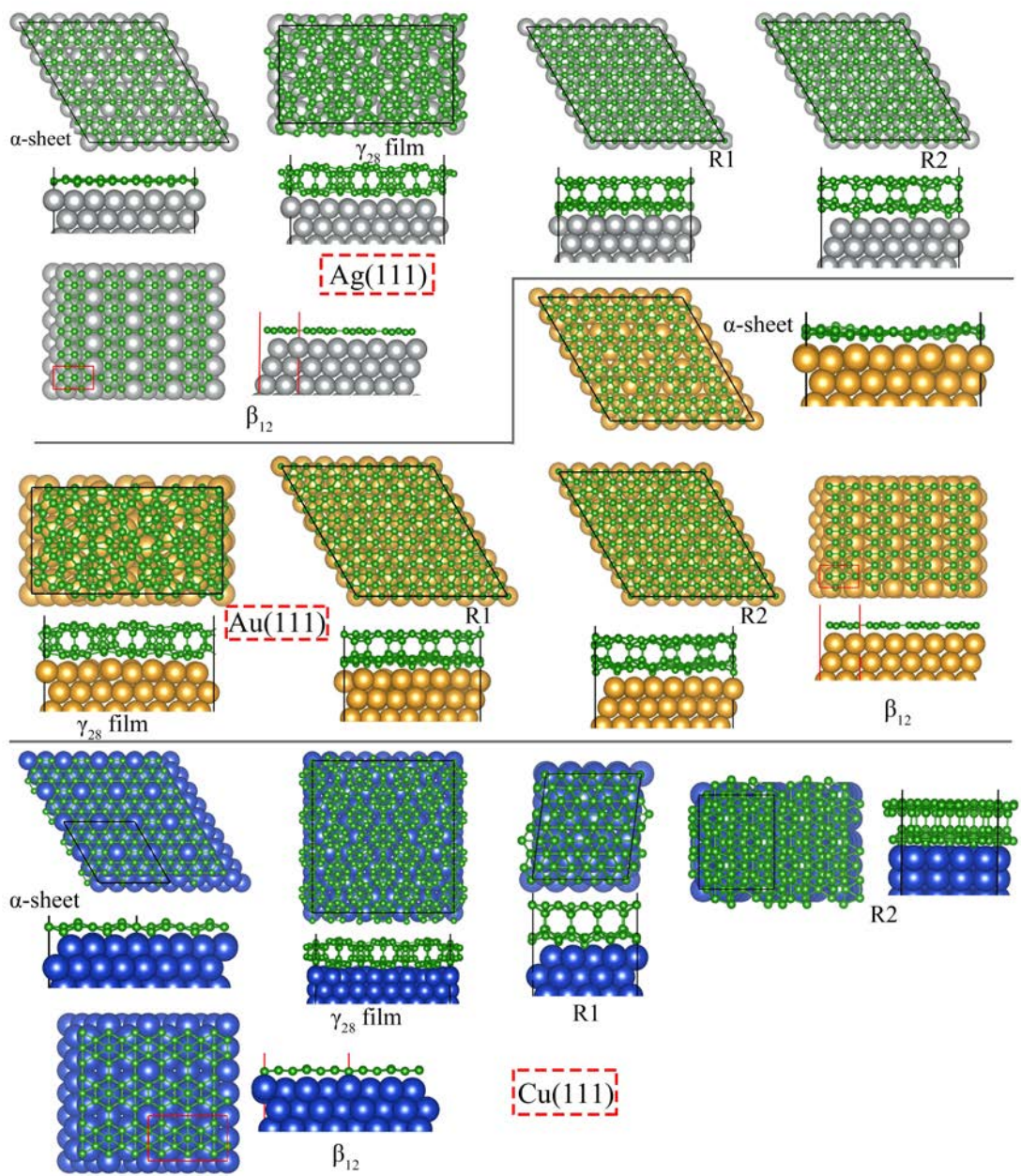

Figure S9. The atomic structures(top view and side view) of the 2D boron allotropes growth on the metal substrates. The solid lines represent the unit cells of the corresponding absorption systems.